\documentclass[12pt]{article}
\usepackage{amssymb,latexsym}
\newtheorem{theo}{Theorem} 
\newtheorem{defi}[theo]{Definition}

\def\beq{\begin{equation}} 
\def\eeq{\end{equation}}
\def\bea{\begin{eqnarray}} 
\def\eea{\end{eqnarray}}
\def\beas{\begin{eqnarray*}} 
\def\eeas{\end{eqnarray*}} 
\def\nn{\nonumber}
\def\lb{[\![} 
\def\rb{]\!]} 
\def\al{\alpha} 
\def\be{\beta} 

\def\de{\delta} 
\def\ep{\epsilon}  
\def\varep{\varepsilon}

\def\De{\Delta}

\def\Om{\Omega}
 
\def\Z{\mathbb{Z}} 
\def\N{\mathbb{N}} 

\begin{document}
\begin{center} 
{\Large \bf 
Quasiboson representations of $sl(n+1)$ and 
generalized quantum statistics}\\[5mm] 
{\bf T.D.\ Palev}\\ 
Institute for Nuclear Research and Nuclear Energy,\\ 
Boul.\ Tsarigradsko Chaussee 72, 1784 Sofia, Bulgaria\\[2mm] 
{\bf J.\ Van der Jeugt\footnote{Research Associate of the
Fund for Scientific Research -- Flanders (Belgium).}}\\ 
Department of Applied Mathematics and Computer Science, 
University of Ghent,\\
Krijgslaan 281-S9, B-9000 Gent, Belgium. 
\end{center}

\vskip 5mm
\begin{abstract}
Generalized quantum statistics will be presented in the context of
representation theory of Lie (super)algebras.
This approach provides a natural mathematical framework, as is
illustrated by the relation between para-Bose and para-Fermi operators 
and Lie (super)algebras of type $B$. 
Inspired by this relation, $A$-statistics
is introduced, arising from representation theory of the Lie algebra
$A_n$. The Fock representations for $A_n=sl(n+1)$ provide microscopic
descriptions of particular kinds of exclusion statistics, which may be
called quasi-Bose statistics. It is indicated that $A$-statistics appears
to be the natural statistics for certain lattice models in condensed
matter physics. 
\end{abstract} 
\vskip 2mm

\section{Introduction}

During the last two decades quantum statistics became a field
of increasing interest among field theorists and condensed matter
theorists.  Various new statistics were suggested, leading to
generalizations or deviations from some of the first principles
in quantum physics, such as the Heisenberg commutation relations,
the Pauli exclusion principle and the commutativity of
space-time.

Some of the generalizations of quantum statistics arose because
of new developments in mathematics.
For example, the theory of quantum groups led to the introduction of
deformed Bose creation and annihilation operators
(CAOs)~\cite{PW}. 
In a similar spirit, multi-mode $q$-deformed CAOs were
introduced, leading to the so-called quon algebra~\cite{Greenberg} and the
related quon statistics. 
In the context of non-commutative geometry~\cite{Connes},
deformations of the Heisenberg commutation relations have been
studied too.

In condensed matter physics the discovery of the fractional quantum
Hall effect in two-dimensional electron gasses led to the theoretical
study of anyons (``particles" with fractional statistics) in
two-dimensional systems~\cite{Wilczek}.

A further breakthrough in the area of quantum statistics was
marked with the paper of Haldane~\cite{Haldane}, who proposed a
generalized version of the Pauli exclusion principle.
This new statistics is now referred to as (fractional) exclusion
statistics.
The validity of exclusion statistics has been tested on
several examples, e.g.\ spinon excitations in a spin-${1\over 2}$
quantum antiferromagnetic chain, anyon gas and anyons in a strong
magnetic field, particles in a one-dimensional Luttinger liquid, $\ldots$.
Also in conformal field theories exclusion statistics is known to play a
role~\cite{Schoutens}.

Our own approach to generalized quantum statistics is inspired by
Wigner's ideas for noncanonical quantum systems~\cite{Wigner}
and the generalizations of Green in quantum field theory,
which led to the introduction of
para-Bose and para-Fermi statistics~\cite{Green}.
These statistics are now known to belong to the class of Lie
(super)algebras of type $B$.
Motivated by this relation, we introduce the more general concept
of statistics belonging to a Lie (super)algebra.
In particular, quantum statistics belonging to the Lie algebra
$A_n$ is discussed here, and referred to as $A$-statistics.
Some of the basic properties of $A$-statistics are summarized.
It is shown that $A$-statistics has an interpretation as exclusion
statistics.
Furthermore, we show that $A$-statistics can be considered as a
finite-dimensional approximation to Bose statistics.
For this reason, it can also be referred to as quasi-Bose statistics, 
and the corresponding CAOs are called
quasibosons or quasi-Bose operators. The quasi-Bose operators can be
expressed as functions of the ordinary Bose operators, a fact which is
related to the Holstein-Primakoff realization of $sl(n+1)$. Finally, we
present an example from condensed matter physics where
the quasibosons play a role in the solution of the problem.

\section{Para-Bose and para-Fermi statistics}

In 1950, Wigner~\cite{Wigner}
showed that canonical quantum statistics can
be generalized in a logically consistent way.
In particular, abandoning the canonical commutation relation
$[p,q]=-i$, one can search for all operators $q$ and $p$ such that
the ``classical'' equations of motion $\dot p=-q$ and $\dot q =p$
are compatible (equivalent) with the Heisenberg equations
$\dot p =-i[p,H]$, $\dot q= -i[q,H]$, where $H$ is the Hamiltonian
of the system. Wigner solved this problem
for the case of a
one-dimensional oscillator ($H={1\over 2}(p^2+q^2)$), and showed that
apart from the canonical solution there are infinitely many other
solutions. For a $n$-dimensional oscillator, the compatibility conditions
read~: 
\beq 
\sum_{i=1}^n [ \{a_i^+, a_i^-\}, a_k^{\pm}] = \pm 2 a_k^{\pm}.
\label{CC} 
\eeq 
Herein, $a_k^{\pm}={1\over\sqrt{2}}(q_k\mp i p_k)$.
The problem is thus reduced
to finding all operators satisfying~(\ref{CC}). Or, reformulated in a
different way~: find all (Hermitian) representations of the associative
algebra generated by the elements $a_k^\pm$ subject to the
relations~(\ref{CC}).

One familiar solution is of course the canonical solution,
since the ordinary Bose CAOs do
satisfy~(\ref{CC}). The corresponding representations space is
the familiar boson Fock space.

Another set of solutions are provided by the para-Bose operators.
These operators $b_i^\pm$ satisfy the triple relations
\beq
[\{ b_i^\xi, b_j^\eta\}, b_k^\epsilon ] =
2\epsilon \delta_{jk}\delta_{\epsilon,-\eta} b_i^\xi +
2\epsilon \delta_{ik}\delta_{\epsilon,-\xi} b_j^\eta,
\label{pB}
\eeq
where $i,j,k=1,2,\ldots,n$ and $\xi,\eta,\epsilon=\pm,\pm 1$.
In quantum field theory, these operators were introduced by
Green~\cite{Green} as a possible generalization of 
statistics of integer-spin fields.

Green also generalized Fermi statistics to para-Fermi statistics
by introducing the para-Fermi operators $f_i^\pm$, satisfying
\beq
[ [ f_i^\xi, f_j^\eta], f_k^\epsilon ] =
2 \delta_{jk}\delta_{\epsilon,-\eta} f_i^\xi -
2 \delta_{ik}\delta_{\epsilon,-\xi} f_j^\eta.
\label{pF}
\eeq

The para-Bose (resp.\ para-Fermi) algebra can be defined as the
associative algebra generated by the elements $b_i^\pm$ (resp.\
$f_i^\pm$) subject to the relations~(\ref{pB}) (resp.~(\ref{pF})).
It is then of importance to classify all irreducible (Hermitian Fock)
representations of the para-Bose and para-Fermi algebra. For the
para-Fermi algebra, it was realized by Kamefuchi and
Takahashi~\cite{Kamefuchi}, and independently by Ryan and
Sudarshan~\cite{Ryan}, that the linear envelope of the elements 
\[
f_i^\xi, \ [f_j^\eta, f_k^\epsilon],\qquad (i,j,k=1,\ldots,n,\
\xi,\eta,\epsilon=\pm) 
\] 
is the orthogonal Lie algebra $B_n=so(2n+1)$.
Thus the representation theory of the para-Fermi algebra
reduces to a problem of representation theory of the Lie algebra $B_n$.

In a similar way, Ganchev and Palev~\cite{Ganchev}
showed that the linear envelope
of the elements
\[
b_i^\xi, \ \{b_j^\eta, b_k^\epsilon\},\qquad
(i,j,k=1,\ldots,n,\ \xi,\eta,\epsilon=\pm)
\]
is the orthosymplectic Lie superalgebra $B(0,n)=osp(1/2n)$,
thereby reducing the representation
theory of the para-Bose algebra to a problem in representation
theory of Lie superalgebras.

\section{Statistics related to a Lie (super)algebra}

As para-Bose statistics is belonging to the class of Lie superalgebras
$B(0,n)$ and para-Fermi statistics to the class of Lie algebras $B_n$, one
can wonder what type of statistics is belonging to the other classes of
Lie (super)algebras.

Let $\cal A$ be a Lie (super)algebra, with bracket $\lb\,,\,\rb$
(which could stand for a commutator $[\,,\,]$ or an anti-commutator
$\{\,,\,\}$).
\begin{defi}
A set of root vectors $a_1^\xi,\ldots,a_n^\xi$ ($\xi=\pm$)
are creation and annihilation operators for $\cal A$ if
\begin{itemize}
\item ${\cal A}$ is equal to the linear envelope of
\[
a_i^\xi, \ \lb a_j^\eta, a_k^\epsilon\rb ,\qquad
(i,j,k=1,\ldots,n,\ \xi,\eta,\epsilon=\pm);
\]
\item
$a^+_i$ (resp.\ $a_i^-$) are negative (resp.\ positive) root vectors.
\end{itemize} 
\end{defi} 
At the same time, one should also select the
representations of the algebra that could play a role in physics by making
the following requirements. 
\begin{defi} 
The ${\cal A}$-module $W$ is a
Fock space if it is a Hilbert space such that 
\begin{itemize} 
\item
$(a_i^+)^\dagger = a_i^-$ (Hermiticity); 
\item there exists a vector
$|0\rangle \in W$ such that $a_i^-|0\rangle=0$ for all $i$ (existence of
vacuum); 
\item $W$ is spanned on vectors of the type $a_{i_1}^+ a_{i_2}^+
\cdots a_{i_m}^+|0\rangle$ (irreducibility). 
\end{itemize} 
\end{defi}
These definitions are a straightforward generalization of the properties
that hold for para-Bose and para-Fermi operators and their
representations.

One can now turn its attention to other classes of Lie (super)algebras. Of
particular interest are the statistics related to Lie (super)algebras of
type $A$. For ${\cal A}=A_n=sl(n+1)$, the statistics will be called
$A$-statistics~\cite{Palev77,Palev00}; its related CAOs
will be called $A$-CAOs. 
For ${\cal A}=A(0,n)=sl(1/n)$, the statistics will be called
$A$-superstatistics~\cite{Palev80}; its related CAOs
will be called $A$-super CAOs.
The rest of this paper is
devoted to $A$-statistics and its properties; for a more elaborate
treatment of $A$-statistics, including proofs, see Ref.~\cite{Palev00}.

\section{Properties of $A$-statistics}

We begin this section by recalling some well-know facts about the
Lie algebras $gl(n+1)$ and $sl(n+1$). A basis for $gl(n+1)$ is given
by the elements $e_{ij}$, $i,j=0,1,\ldots,n$, satisfying the relations 
\[
[e_{ij},e_{kl}]=\de_{jk}e_{il}-\de_{il}e_{kj}. 
\] 
The Cartan subalgebra
$H$ of $gl(n+1)$ is spanned by the elements $h_i=e_{ii}$
($i=0,1,\ldots,n$), and let $\ep_i$ be the dual basis of the dual space
$H^*$. Then every element $e_{ij}$ ($i\ne j$) is a root vector with
corresponding root $\ep_i-\ep_j$. The positive roots consist of
$\ep_i-\ep_j$ with $i<j$, and the negative roots consist of $\ep_i-\ep_j$
with $i>j$.

The CAOs for the Lie algebra
$sl(n+1)\subset gl(n+1)$ are chosen as follows~:
\[
a_i^+= e_{i,0}, \qquad a_i^-=e_{0,i},\qquad (i=1,2,\ldots,n).
\]
{}From the defining relations, it is 
easy to verify that
the linear envelope of
\[
a_i^\xi, \ [ a_j^\eta, a_k^\epsilon] ,\qquad
(i,j,k=1,\ldots,n,\ \xi,\eta,\epsilon=\pm),
\]
is indeed $sl(n+1)=A_n$.
Furthermore, it can be shown that the commutation relations of
$A_n$ are completely equivalent with the following set of relations~: 
\bea
&& [a_i^+,a_j^+]= [a_i^-,a_j^-]= 0, \nn\\ &&
[[a_i^+,a_j^-],a_k^+]=\de_{jk} a_i^+ + \de_{ij} a_k^+ , \label{ACC}\\ &&
[[a_i^+,a_j^-],a_k^-]= -\de_{ik} a_j^- - \de_{ij} a_k^-. \nn 
\eea 
These
relations are similar to the para-Bose or para-Fermi relations in the
sense that they also include triple relations. On the other hand, the
actual structure of the relations is quite different and they will also
lead to a different type of statistics. We shall refer to~(\ref{ACC}) as
the relations of $A$-statistics, and call the corresponding CAOs
$A$-operators. The description of $A_n$ by means of
the generators $a_i^\pm$ and the relations~(\ref{ACC}) was already given
in a paper by Jacobson~\cite{Jacobson}; therefore, the elements $a_i^\pm$
could also be referred to as Jacobson generators of $A_n$.

The Fock spaces of $A$-statistics can be classified using representation
theory of $A_n$ and the actual commutation relations between the
generators. The following characterization holds~: 
\begin{theo} 
$W$ is a Fock space of $A_n$ if and only if there exists a
positive integer $p$ such that 
\[ 
a_i^- |0\rangle = 0, \qquad\hbox{and}\qquad a_i^- a_j^+
|0\rangle = p \de_{ij} |0\rangle, 
\] 
for $i,j=1,2,\ldots,n$. The Fock
space is uniquely characterized by the number $p$. 
\end{theo} 
Because of
the resemblance with Green's definition of ``parastatistics of order
$p$'', we shall also refer here to $p$ as the order of the $A$-statistics.

Let $W_p$ be the Fock space characterized by $p$. The following
summarizes a number of properties of $W_p$~:
\begin{itemize}
\item
$W_p$ is spanned on vectors 
\[
(a_1^+)^{l_1} (a_2^+)^{l_2} \cdots (a_n^+)^{l_n} |0\rangle
\]
with $l_i=0,1,\ldots$.
Any such vector is nonzero if and only if
\[
l_1+l_2+\cdots +l_n \leq p.
\]
\item
$W_p$ is a finite dimensional highest weight module of $A_n$ with
highest weight $(p,0,\ldots,0)$ in the $\ep_i$-basis.
\item
The scalar product on $W_p$ is completely characterized by
\beas
&& \langle 0 | 0\rangle =1,\\
&& \langle v | a_i^+ w \rangle = \langle a_i^- v | w \rangle 
, \qquad \forall v,w \in W_p.
\eeas
Thus an orthonormal basis for $W_p$ is given by the vectors
\beq
|p;l_1,\ldots,l_n\rangle=\sqrt{(p-\sum_{j=1}^n l_j )!\over p!}
{(a_1^+)^{l_1}\ldots (a_n^+)^{l_n}\over \sqrt{l_1!l_2!\ldots
l_n!}} |0\rangle,\quad l_1+l_2+\ldots+l_n\leq p .
\label{orthbasis}
\eeq
\item
The transformation of the basis~(\ref{orthbasis}) under the action
of the CAOs reads~:
\bea
&& a_i^+|p;l_1,\ldots ,l_i,\ldots,l_n\rangle =
  \sqrt{(l_i+1)(p-\sum_{j=1}^n l_j  )}\
  |p;l_1\ldots,l_i+1,\ldots,l_n\rangle,
\label{action+}\\
&& a_i^-|p;l_1,\ldots,l_i,\ldots,l_n\rangle=
  \sqrt{l_i(p-\sum_{j=1}^n l_j +1  )}\
  |p;l_1\ldots,l_i-1,\ldots,l_n\rangle.
\label{action-}
\eea
\end{itemize}
The last equations indicate that the operators $a_i^+$ (resp.\ $a_i^-$)
can be interpreted as creating (resp.\ annihilating) a ``particle on the
$i^{\rm th}$ orbital''. 
This can be further supported by introducing a Hamiltonion 
\[ 
H=\sum_{i=1}^n \varep_i h_i = \sum_{i=1}^n \varep_i
([a_i^+,a_i^-]+h_0). 
\] 
This Hamiltonian satisfies 
\[
[H,a_i^\pm] = \pm \varep_i a_i^\pm 
\]
and therefore the operators $a_i^+$
(resp.\ $a_i^-$) can be interpreted as creating (resp.\ annihilating) a
``particle with energy $\varep_i$''.

Clearly, $H$ belongs to the Cartan subalgebra of
$gl(n+1)$, and not to $sl(n+1)$. Therefore, it cannot be written as a
function of the CAOs only. However, within
a fixed irreducible representation $W_p$, $H$ can be represented as
follows~: 
\[ 
H={1\over{n+1}}\sum_{i=1}^n \varep_i\Bigl(p+n[a_i^+,a_i^-]-
\sum_{k\ne i=1}^n [a_k^+,a_k^-]\Bigr). 
\]

\section{The Pauli principle for $A$-statistics}

Let us consider, as an example, $A$-statistics of order $p=4$
with $n=6$ orbitals, 
corresponding to 6 different energy levels.
{}From~(\ref{orthbasis}), it follows that there is no restriction
on the number of particles to be accommodated on a certain 
orbital or, which is the same, on a certain energy level, 
as long as the total number of particles in 
any configuration does not exceed $p$.
Hence, the following three states or configurations are allowed
(the orbitals, i.e., the energy levels, are represented by lines, and
the particles by dots)~:
\[
\vbox{
\unitlength=1mm
\special{em:linewidth 0.4pt}
\linethickness{0.4pt}
\begin{picture}(80.00,145.00)
\put(10.00,145.00){\line(1,0){10.00}}
\put(40.00,145.00){\line(1,0){10.00}}
\put(70.00,145.00){\line(1,0){10.00}}
\put(10.00,140.00){\line(1,0){10.00}}
\put(40.00,140.00){\line(1,0){10.00}}
\put(70.00,140.00){\line(1,0){10.00}}
\put(10.00,135.00){\line(1,0){10.00}}
\put(40.00,135.00){\line(1,0){10.00}}
\put(70.00,135.00){\line(1,0){10.00}}
\put(10.00,130.00){\line(1,0){10.00}}
\put(40.00,130.00){\line(1,0){10.00}}
\put(70.00,130.00){\line(1,0){10.00}}
\put(10.00,125.00){\line(1,0){10.00}}
\put(40.00,125.00){\line(1,0){10.00}}
\put(70.00,125.00){\line(1,0){10.00}}
\put(10.00,120.00){\line(1,0){10.00}}
\put(40.00,120.00){\line(1,0){10.00}}
\put(70.00,120.00){\line(1,0){10.00}}
\put(15,130){\circle*{1.5}}
\put(15,125){\circle*{1.5}}
\put(45,140){\circle*{1.5}}
\put(43,130){\circle*{1.5}}
\put(47,130){\circle*{1.5}}
\put(45,125){\circle*{1.5}}
\put(72,135){\circle*{1.5}}
\put(75,135){\circle*{1.5}}
\put(78,135){\circle*{1.5}}
\put(75,120){\circle*{1.5}}
\end{picture}
\vskip -122mm
}
\]
Note that the last two configurations are already ``saturated'' in
the sense that no more particles can be added, since the total
number of particles is already equal to $p=4$.
The following two configurations correspond to
forbidden states~:
\[
\vbox{
\unitlength=1mm
\special{em:linewidth 0.4pt}
\linethickness{0.4pt}
\begin{picture}(50.00,145.00)
\put(10.00,145.00){\line(1,0){10.00}}
\put(40.00,145.00){\line(1,0){10.00}}
\put(10.00,140.00){\line(1,0){10.00}}
\put(40.00,140.00){\line(1,0){10.00}}
\put(10.00,135.00){\line(1,0){10.00}}
\put(40.00,135.00){\line(1,0){10.00}}
\put(10.00,130.00){\line(1,0){10.00}}
\put(40.00,130.00){\line(1,0){10.00}}
\put(10.00,125.00){\line(1,0){10.00}}
\put(40.00,125.00){\line(1,0){10.00}}
\put(10.00,120.00){\line(1,0){10.00}}
\put(40.00,120.00){\line(1,0){10.00}}
\put(15,120){\circle*{1.5}}
\put(15,125){\circle*{1.5}}
\put(15,130){\circle*{1.5}}
\put(15,135){\circle*{1.5}}
\put(15,140){\circle*{1.5}}
\put(45,140){\circle*{1.5}}
\put(47,130){\circle*{1.5}}
\put(43,130){\circle*{1.5}}
\put(47,125){\circle*{1.5}}
\put(43,125){\circle*{1.5}}
\end{picture}
\vskip -122mm
}
\]
Both of these states are not allowed since the total number of
particles in the configuration exceeds $p=4$.

This example clearly illustrates the accommodation properties of
$A$-statistics of order $p$. Because of this ``exclusion principle'',
$A$-statistics can be shown to be a special case of exclusion statistics.
Haldane introduced (fractional) exclusion statistics~\cite{Haldane} by
means of the relation $ \De d = -g \De N$ (for one kind of particles),
where $\De d$ is the change in
dimension of a single-particle Hilbert space, $\De N$ is the allowed
increase of the number of particles, and $g$ is the constant
characterizing the exclusion statistics.
One possible ``integral'' solution of this relation 
can be written as~\cite{Wu} 
\beq 
d(N)=n-g\cdot(N-1). 
\label{ES} 
\eeq 
This should be interpreted as follows~: 
let $n$ be the total number of orbitals
that are available for the first particle, and suppose $N-1$ particles are
already accommodated in the configuration, then $d(N)$ expresses the
dimension of the single-particle space for the $N^{\rm th}$ particle (or
the number of orbitals where the $N^{\rm th}$ particle can be ``loaded'').
Bose statistics has $g=0$, and Fermi statistics has $g=1$.

If one accepts the natural assumption that~(\ref{ES}) 
should hold for all {\em admissible}
values of $N$, i.e.\ one does  
not require~(\ref{ES}) to be applicable for values of $N$
which the system cannot accommodate, then
$A$-statistics is a particular case of exclusion statistics, also with
$g=0$. $A$-statistics is similar to Bose statistics in the sense that
there is no restriction on the number of particles on an orbital. The main
difference comes from the fact that the total configuration should contain
no more than $p$ particles. For this reason, we shall refer to
$A$-statistics also as quasi-Bose statistics.

\section{Quasi-Bose creation and annihilation operators}

Introducing new (representation dependent) CAOs by
\beq
B(p)_i^\pm = a_i^\pm / \sqrt{p}, \qquad i=1,\ldots,n,
 \qquad p\in \N,
\label{quasiboson}
\eeq
the transformation formulas (\ref{action+}) and (\ref{action-}) become
\bea && B(p)_i^+|p;l_1,\ldots,l_n\rangle=
  \sqrt{(l_i+1)(1-{{\sum_{k=1}^n l_k}\over p})}\
|p;l_1,\ldots,l_i+1,\ldots,l_n\rangle, \label{actionB+}\\
&&B(p)_i^-|p;l_1,\ldots,l_n\rangle=
  \sqrt{l_i(1+{{1-\sum_{k=1}^n l_k}\over p})}\
|p;l_1,\ldots,l_i-1,\ldots,l_n\rangle \label{actionB-}.
\eea
Suppose that the total number of particles in the configuration
is much less than the order of the statistics, i.e.\
$ l_1+l_2+\ldots+l_n \ll p$. In this approximation one obtains~:
\bea
&& B(p)_i^+|p;l_1,\ldots,l_n\rangle \simeq
  \sqrt{l_i+1}\;|p;l_1,\ldots,l_i+1,\ldots,l_n\rangle,
\label{actionB+appr}\\
&& B(p)_i^-|p;l_1,\ldots,l_n\rangle \simeq
  \sqrt{l_i}\; |p;l_1\ldots,l_i-1,\ldots,l_n\rangle,
\label{actionB-appr}
\eea
which yields an approximation to the Bose commutation relations~:
\beas
&& [B(p)_i^+,B(p)_j^+]=[B(p)_i^-,B(p)_j^-]=0,\qquad ({\rm exact})\\
&& [B(p)_i^-,B(p)_j^+]\simeq
\de_{ij}, \qquad {\rm if}\  l_1+l_2+\ldots+l_n \ll p.
\eeas
This is the main reason to refer to the operators $B(p)_i^\pm$ as
quasi-Bose CAOs, or quasibosons.

The above described approximation can be formulated in a mathematically
correct way, by introducing the Hilbert space $\ell_2(\Z_+^n)$ with
orthonormal basis $|l_1,\ldots, l_n\rangle$ ($l_i\in\Z_+$). On this
Hilbert space, quasi-Bose operators and Bose operators 
are linear operators defined by means of the action ($p\in\N$) 
\beas 
&& B(p)_i^+|l_1,\ldots,l_n\rangle=
  \sqrt{(l_i+1)(1-{{\sum_{k=1}^n l_k}\over p})}\
|l_1,\ldots,l_i+1,\ldots,l_n\rangle, \hbox{ for }\sum_k l_k\leq p,\\
&&B(p)_i^-|l_1,\ldots,l_n\rangle=
  \sqrt{l_i(1+{{1-\sum_{k=1}^n l_k}\over p})}\
|l_1,\ldots,l_i-1,\ldots,l_n\rangle, \hbox{ for }\sum_k l_k\leq p,\\
&&B(p)_i^\pm |l_1,\ldots,l_n\rangle= 0,
\hbox{ for }\sum_k l_k > p,\\
&& B_i^+|l_1,\ldots,l_n\rangle=
\sqrt{l_i+1}\;|l_1,\ldots,l_i+1,\ldots,l_n\rangle, \\ &&
B_i^-|l_1,\ldots,l_n\rangle=
  \sqrt{l_i}\; |l_1\ldots,l_i-1,\ldots,l_n\rangle.
\eeas
These operators are shown to be continuous operators over 
an appropriate dense subspace $\Om$ of the Hilbert space. 
Then a certain strong topology can be introduced on the set of such 
operators so that $\lim_{p\rightarrow \infty} B(p)_i^\pm = B_i^\pm$,
showing that the quasi-Bose operators are a ``finite-dimensional
approximation'' to the ordinary Bose operators~\cite{Palev00}.

Introducing, as usual, number operators $N_i=B_i^+B_i^-$ with
action $N_i |l_1,\ldots,l_n\rangle= l_i |l_1,\ldots,l_n\rangle$,
one derives from the above action that the quasi-Bose operators
can be realized as follows (that is, for their action on
states with $\sum_k l_k\leq p$)~:
\beq
B(p)_i^+ = B_i^+ \sqrt{1-{\sum_k B_k^+ B_k^- \over p}},\qquad
B(p)_i^- = \sqrt{1-{\sum_k B_k^+ B_k^- \over p}}\, B_i^-.
\label{BpB}
\eeq
This ``bosonisation'' of the quasi-Bose operators can be shown
to be related to the Holstein-Primakoff realization~\cite{HP} of
$gl(n+1)$~\cite{Palev00}.

\section{Quasi-Bose operators in physical model}

The quasi-Bose operators introduced here should prove to be useful
in physical boson models where finite-dimensional state spaces are
required. As an example, consider a two-leg spin-$1/2$ quantum
Heisenberg ladder~\cite{Gopalan} with Hamiltonian
\[
{\hat H}=\sum_{i} (J{\bf{\hat S}}_{i}^+ {\bf{\hat S}}_{i+1}^+ +
J{\bf{\hat S}}_{i}^- {\bf{\hat S}_{i+1}}^- + J_\bot{\bf{\hat
S}}_{i}^+ {\bf{\hat S}}_{i}^-).
\]
Herein, ${\bf{\hat S}}_{i}^\pm\equiv
({\hat S}_{1 i}^\pm,{\hat S}_{2 i}^\pm,{\hat S}_{3 i}^\pm)$
are two commuting spin-$1/2$ vector operators ``sitting" on site $i$ of
the chain 
$\pm$ and the Hamiltonian is a scalar with respect to the total
spin operator ${\bf{\hat S}} =\sum_{i}( {\bf{\hat S}}_{i}^+ + {\bf{\hat
S}}_{i}^-)$~: 
\[ 
[{\hat S}_{\alpha i}^\pm,{\hat S}_{\beta i}^\pm]=
i\sum_{\gamma}\epsilon_{\alpha \beta \gamma}{\hat S}_{\gamma i}^\pm,\qquad
[{\hat S}_{\alpha i}^+,{\hat S}_{\beta j}^-]=0,\qquad [{\hat H},{\bf {\hat
S}}]=0. 
\] 
If $J_\bot \gg J$ 
(disordered phase) the state of the system is
well described with the bond operator representation of spin
operators~\cite{Chubukov, Sachdev}. 
Following~\cite{Sachdev}, the spin 
operators in~\cite{Gopalan} were expressed in terms of four pairs of bosons
per site subject to an additional constraint.  
In~\cite{Sushkov} ${\bf{\hat S}}_{i}^\pm$ were realized via three pairs
of Bose operators $B_{\al i}^\pm$ per site:
\beq 
{\hat S}_{\alpha i}^\pm={1\over 2}(\pm B_{\alpha i}^- \pm
B_{\alpha i}^+ -i\epsilon_{\alpha \beta \gamma}B_{\beta i}^+B_{\gamma
i}^-),~~ \alpha, \beta, \gamma=1,2,3. 
\label{bosonisation} 
\eeq 
The introduction of such Bose
operators is convenient for the solution of the model, but also creates
certain problems. For instance, the local state space at site $i$ becomes
infinite dimensional (whereas it should be only 4-dimensional with the
given spin-$1/2$ states). One way to overcome this problem is to
introduce ``by hands'' an infinite on-site repulsion~\cite{Sushkov}.
Another approach is to use the bond operator representation of
Chubukov~\cite{Chubukov}. In this case the Bose operators $B_{\al
i}^\pm$ in~(\ref{bosonisation}) are replaced
by new operators $b_{\al i}^\pm$ as follows~:
\[ 
B_{\al i}^+ \rightarrow b_{\al i}^+ = B_{\al i}^+ \sqrt{ 1-
\sum_{\be=1}^3 B_{\be i}^+ B_{\be i}^-}, \qquad B_{\al i}^- \rightarrow
b_{\al i}^- = \sqrt{ 1- \sum_{\be=1}^3 B_{\be i}^+ B_{\be i}^-} B_{\al
i}^- . 
\] 
A close look at these expressions reveals that one is actually
using the quasi-Bose operators of order $p=1$ in their Bose
realization~(\ref{BpB}). Replacing the Bose operators with $p=1$
quasi-Bose operators throughout the model, one obtains directly the
physical state space and the correct expansions for the spin operators and
the Hamiltonian. This shows the practical use of the quasi-Bose operators
in this model. From the intrinsic properties of quasibosons, it is clear
that the quasi-Bose operators should be useful in other physical models as
well.

\end{document}